\documentclass
[preprintnumbers,superscriptaddress,pra,showpacs,showkeys]{revtex4}%
\usepackage[colorlinks=true,linkcolor=blue]{hyperref}
\usepackage{amssymb}
\usepackage{amsfonts}
\usepackage{amsmath}
\usepackage{graphicx}%
\providecommand{\U}[1]{\protect\rule{.1in}{.1in}}

\let\stdsection\section
\renewcommand\section{\nopagebreak\stdsection}
\begin{document}
\title[Geometric momentum on two-dimensional sphere]{Distribution of $xp$ in some molecular rotational states}
\author{Q. H. Liu}
\email{quanhuiliu@gmail.com}
\affiliation{School for Theoretical Physics, and Department of Applied Physics, Hunan
University, Changsha, 410082, China}
\date{\today}

\begin{abstract}
Developing the analysis of the distribution of the so-called posmom $xp$ to
the spherical harmonics that represents some molecular rotational states for
such as diatomic molecules and spherical cage molecules, we obtain posmometry
(introduced recently by Y. A. Bernard and P. M. W. Gill, Posmom: The
Unobserved Observable, J. Phys. Chem. Lett. 1\textbf{(}2010\textbf{)}1254) of
the spherical harmonics and demonstrate that it bears a striking resemblance
to the momentum distributions of the stationary states for a one-dimensional
simple harmonic oscillator.

\end{abstract}

\maketitle

\section{Introduction}

The quantum mechanics operator of type $xp$ has attracted much attention since
the invention of quantum mechanics. In the quantum measurement theory, a
frequently quoted pointer's Hamiltonian that was given by John von Neumann is
taken as $H=-\omega xp$ where $\omega$ is some constant, and $x$ is a pointer
coordinate and $p$ is the pointer momentum operator.
\cite{neumann,weinberg,weakM,xmzhu} Recently, Gill et al. intensively explore
the particle's position-momentum dot product $\mathbf{r}\cdot\mathbf{p}$, or
\textit{posmom} as they called, and establish a \textit{posmometry} (the
distribution density of the posmom) for some atomic and molecular systems.
\cite{posmometry1,posmometry2} They consider that the posmom density provides
unique insight into types of trajectories electrons may follow, complementing
existing spectroscopic techniques that provide information about where
electrons are (X-ray crystallography) or where they go (Compton spectroscopy).
\cite{posmometry1,posmometry2} The \textit{posmom }operator in one component
of the three dimensional Cartesian coordinates, e.g., $Q_{i}\equiv
(x_{i}p_{x_{i}}+p_{x_{i}}x_{i})/2,(i=1,2,3)$, is an essentially self-adjoint
operator, \cite{winter,milburn} and has already been studied as part of a
larger space-time conformal transformation in certain non-relativistic quantum
mechanical problems by de Alfaro et al, \cite{diluton} Jackiw. \cite{jackiw}
Currently, the operator $Q_{i}$ arouses broad interest
\cite{milburn,broad,berry} due to the pioneer work of Berry and Keating who
take this operator $Q_{x}$ to be the Hamiltonian of a system and demonstrate
that there are possible connections between the Riemann conjecture and
eigenfunctions of the operator. \cite{berry} In this paper, we put the
operator $Q_{i}$ on a two-dimensional spherical surface $S^{2}$ and work out
the distribution of $Q_{z}$ for some molecular rotational states. Precisely,
we hope to give the posmometry for spherical harmonics that describes the
rotational states for some molecules, such as diatomic molecules, spherical
cage molecule $C_{60}$\ or $Au_{32}$, etc., of which the vibrational and
rotational motions are weakly coupled so that the rotational modes can be
independently treated.

Note that with embedding $S^{2}$ in the three-dimensional flat space $R^{3}$,
there are three operators $Q_{i}$ $(i=1,2,3)$ that are respectively defined
along three axes of coordinate respectively. These three operators are
equivalent to each other upon axis rotations or relabeling. \cite{sunhr}
Moreover, it is easy to show that there are three pairs of complete set of
commuting operators $[Q_{i},L_{i}]=0$ ($i=1,2,3$) each of which offers
complete description of the states on $S^{2}$, but they are also equivalent to
each other upon axis rotations or relabeling, \cite{sunhr} where $L_{i}$ is
the $i$th components of orbital angular momentum.

This paper is organized as what follows. In section II, we give the complete
set of eigenfunctions $\psi_{\lambda_{z}}$ of the ($Q_{z}$, $L_{z}$). In
section III, we show how to expand spherical harmonics on the bases
$\psi_{\lambda_{z}}$. The section IV briefly concludes this study.

\section{Complete set of eigenfunctions constructed from simultaneous
eigenfunctions of two commuting operators ($Q_{z}$, $L_{z}$)}

For $S^{2}$ of fixed radius $r$, parametrized by,
\begin{equation}
x=r\sin\theta\cos\varphi,\text{ }y=r\sin\theta\sin\varphi,\text{ }%
z=r\cos\theta,
\end{equation}
the preliminary form of the momentum is $\mathbf{p}^{\prime}=-i\hbar
\nabla_{s^{2}}$ ($\nabla_{s^{2}}$ is the gradient operator on $S^{2}$
\cite{geometry}), defined by,%

\begin{equation}
\mathbf{p}^{\prime}=-i\hbar\frac{1}{r}\left(  \cos\theta\cos\varphi
\frac{\partial}{\partial\theta}-\frac{\sin\varphi}{\sin\theta}\frac{\partial
}{\partial\varphi},\cos\theta\sin\varphi\frac{\partial}{\partial\theta}%
+\frac{\cos\varphi}{\sin\theta}\frac{\partial}{\partial\varphi},-\sin
\theta\frac{\partial}{\partial\theta}-\cos\theta\right)  .
\end{equation}
We construct the self-adjoint operators $Q_{i}$ in the following symmetric way
with noting $p_{i}^{\prime}x_{i}\neq(x_{i}p_{i}^{\prime})^{\dagger}$ because
of the presence of the nontrivial metric\ factor $1/\sin\theta$,%
\begin{equation}
Q_{i}=\frac{1}{2}\left\{  \frac{1}{2}\left[  x_{i}p_{i}^{\prime}+(x_{i}%
p_{i}^{\prime})^{\dagger}\right]  +\frac{1}{2}\left[  p_{i}^{\prime}%
x_{i}+(p_{i}^{\prime}x_{i})^{\dagger}\right]  \right\}  .
\end{equation}
The results turn out to be,%
\begin{equation}
Q_{i}=\frac{1}{2}\left(  x_{i}p_{i}+p_{i}x_{i}\right)  ,
\end{equation}
where the momenta $p_{i}$ ($i=1,2,3$) are the so-called geometric momentum
$\mathbf{p}=-i\hbar(\nabla_{s^{2}}+M\mathbf{n})$
\cite{liu11,liu13,03,07,10,131,132} on $S^{2}$ where $M$ is the mean curvature
$-1/r$ and $\mathbf{n}$ is the normal vector, which is proposed for a proper
description of momentum in quantum mechanics for a particle constrained on
$S^{2}$, and explicitly we have, \cite{liu11,liu13,03,07,10,131,132}
\begin{align}
p_{x}  &  =-i\hbar\frac{1}{r}(\cos\theta\cos\varphi\frac{\partial}%
{\partial\theta}-\frac{\sin\varphi}{\sin\theta}\frac{\partial}{\partial
\varphi}-\sin\theta\cos\varphi),\\
p_{y}  &  =-i\hbar\frac{1}{r}(\cos\theta\sin\varphi\frac{\partial}%
{\partial\theta}+\frac{\cos\varphi}{\sin\theta}\frac{\partial}{\partial
\varphi}-\sin\theta\sin\varphi),\\
p_{z}  &  =i\hbar\frac{1}{r}(\sin\theta\frac{\partial}{\partial\theta}%
+\cos\theta).
\end{align}

The following properties of the three operators $Q_{i}$ $(i=1,2,3)$ are easily
verifiable: i) Since the geometric momentum $\mathbf{p}$ describes the motion
constrained on the surface $S^{2}$ and there is no motion along the normal
direction $\mathbf{n}$, which in quantum mechanics is expressed by
$\mathbf{x\cdot p+p\cdot x}\equiv2(Q_{x}+Q_{y}+Q_{z})=0$ while it is in
classical mechanics expressed by $\mathbf{x\cdot p}=0$. ii) Three components
($Q_{x},Q_{y},Q_{z}$) are mutually commuting operators so they have
simultaneous eigenfunctions that turn out to be, $\tan^{i(a_{x}+a_{y})}%
\theta\cos^{ia_{x}}\varphi\sin^{ia_{y}}\varphi/(\sin\theta\sqrt{\cos\theta
}\sqrt{\sin2\varphi})$, where the eigenvalues of ($Q_{x},Q_{y},Q_{z}$) are
respectively ($a_{x},a_{y},-(a_{x}+a_{y}))$. iii) There are three pairs of
complete set of commuting observables ($Q_{i},L_{i}$) ($i=x,y,z$), and they
differ from each other upon a matter of relabelling the axes of coordinate.
So, for simplicity, let us study one pair of them ($Q_{z},L_{i}$) in detail.

Operators $Q_{z}$ and $L_{z}$ are essentially independent from each other for
they are dependent on variables $\theta$ and $\varphi$ respectively,
\begin{equation}
Q_{z}=i\hbar(\sin\theta\cos\theta\frac{\partial}{\partial\theta}+\frac{3}%
{2}\cos^{2}\theta-\frac{1}{2}),\text{ }L_{z}=-i\hbar\frac{\partial}%
{\partial\varphi}.
\end{equation}
Thus, each of their simultaneous eigenfunctions $\psi_{\lambda_{z},m}$ is a
product of two eigenfunctions determined by $S_{z}$ and $L_{z}$, respectively.
The eigenfunctions of $L_{z}$ are $\Phi_{m}(\varphi)=e^{im\varphi}/$
$\sqrt{2\pi}$ corresponding to the eigenvalues $m\hbar$. However, finding the
eigenfunctions $\Psi_{\lambda_{z}}$ of $Q_{z}$ in whole full interval of
$\theta\in(0,\pi)$ is in fact a little bit tricky. A direct solution of the
following eigenvalue equation
\begin{equation}
Q_{z}\Psi_{\lambda_{z}}(\theta)=\hbar\lambda_{z}\Psi_{\lambda_{z}}(\theta)
\end{equation}
leads to following two piecewise eigenfunctions,
\begin{equation}
\Psi_{\lambda_{z}}^{I}(\theta)=\frac{1}{\sqrt{2\pi}}\frac{1}{\sin\theta
\sqrt{\cos\theta}}\exp(-i\lambda_{z}\ln\tan\theta)\text{, }\theta\in
(0,\pi/2)\text{, }\lambda_{z}\in(-\infty,\infty)
\end{equation}
and,
\begin{equation}
\Psi_{\lambda_{z}}^{II}(\theta)=\Psi_{\lambda_{z}}^{I}(\pi-\theta),\theta
\in(\pi/2,\pi)\text{.}%
\end{equation}
Curiously, the functions $\Psi_{\lambda_{z}}^{I}(\theta)$ and $\Psi
_{\lambda_{z}}^{II}(\theta)$ are respectively $\delta$-function normalizable
in half intervals $\theta\in(0,\pi/2)$ or $(\pi/2,\pi)$ instead of in full
interval of $\theta\in(0,\pi)$. In order to construct a complete set of the
physically satisfactory eigenstates in the full interval, we need to consider
the parity of the eigenstates with respect to the equator $\theta=\pi/2$. Then
the complete set of the eigenfunctions in whole interval $\theta\in(0,\pi)$
are,
\begin{equation}
\Psi_{\lambda_{z}}(\theta)=\{\Psi_{\lambda_{z}}^{+}(\theta),\Psi_{\lambda_{z}%
}^{-}(\theta)\}
\end{equation}
where $\Psi_{\lambda_{z}}^{+}(\theta)$ and $\Psi_{\lambda_{z}}^{-}(\theta)$
are the even (with superscript $+$) and odd (with superscript $-$) parity
eigenstates, respectively,
\begin{equation}%
\begin{array}
[c]{c}%
\Psi_{\lambda_{z}}^{+}(\theta)=\left(  \Psi_{\lambda_{z}}^{I}(\theta
)+\Psi_{\lambda_{z}}^{II}(\pi-\theta)\right)  /\sqrt{2}\\
\Psi_{\lambda_{z}}^{-}(\theta)=\left(  \Psi_{\lambda_{z}}^{I}(\theta
)-\Psi_{\lambda_{z}}^{II}(\pi-\theta)\right)  /\sqrt{2}%
\end{array}
,\text{ }\theta\in(0,\pi).
\end{equation}
We have therefore in general the simultaneous eigenstates $\psi_{\lambda
_{z},m}(\theta,\varphi)$ of $Q_{z}$ and $L_{z}$,
\begin{equation}
\psi_{\lambda_{z},m}(\theta,\varphi)=\Psi_{\lambda_{z}}(\theta)\Phi
_{m}(\varphi).
\end{equation}
A state $\psi(\theta,\varphi)$ on $S^{2}$ can be expanded in terms of
$\psi_{\lambda_{z},m}(\theta,\varphi)$ in the following way,%
\begin{equation}
\psi(\theta,\varphi)=\sum_{m}\int\left[  c_{m}^{+}(\lambda_{z})\Psi
_{\lambda_{z}}^{+}(\theta)+c_{m}^{-}(\lambda_{z})\Psi_{\lambda_{z}}^{-}%
(\theta)\right]  d\lambda_{z}\Phi_{m}(\varphi),
\end{equation}
where the coefficients $c_{m}^{+}(\lambda_{z})$ and $c_{m}^{-}(\lambda_{z})$
are determined by,%
\begin{equation}
c_{m}^{\pm}(\lambda_{z})=\oint\Phi_{m}^{\ast}(\varphi)\Psi_{\lambda_{z}}%
^{\pm\ast}(\theta)\psi(\theta,\varphi)\sin\theta d\theta d\varphi\text{.}%
\end{equation}

\section{Posmometry for spherical harmonics}

As is well known, the spherical harmonics $Y_{lm}(\theta,\varphi)$ offers a
complete Hilbert space for analyzing any state on $S^{2}$, and also represents
the rotational states for some molecules. Explicitly, the spherical harmonics
$Y_{lm}(\theta,\varphi)$ takes the following form,
\begin{equation}
Y_{lm}(\theta,\varphi)\equiv N_{lm}P_{l}^{m}(\cos\theta)\frac{1}{\sqrt{2\pi}%
}e^{im\varphi},
\end{equation}
with $P_{l}^{m}$ being the associated Legendre polynomial,
\begin{equation}
P_{l}^{m}(x)=\frac{(-1)^{m}}{2^{l}l!}(1-x^{2})^{m/2}\frac{d^{l+m}}{dx^{l+m}%
}(x^{2}-1)^{l} \label{alp}%
\end{equation}
and%
\begin{equation}
N_{lm}=\sqrt{\frac{2l+1}{2}\frac{\left(  l-m\right)  !}{\left(  l+m\right)
!}}.
\end{equation}
In general, the posmometry for spherical harmonics is given by,%
\begin{align}
c_{lm}^{\pm}(\lambda_{z})  &  =\oint\Phi_{m}^{\ast}(\varphi)\Psi_{\lambda_{z}%
}^{\pm\ast}(\theta)Y_{lm^{\prime}}(\theta,\varphi)\sin\theta d\theta
d\varphi\text{.}\nonumber\\
&  =\delta_{m,m^{\prime}}\int\Psi_{\lambda_{z}}^{\pm\ast}(\theta
)N_{lm^{\prime}}P_{l}^{m^{\prime}}(\cos\theta)\sin\theta d\theta. \label{I}%
\end{align}
From this result, we see that the operator $L_{z}$ plays a role of identifying
the $z$-component of the orbital angular momentum represented by the prior
chosen spherical harmonics $Y_{lm}(\theta,\varphi)$, also a choice of the
common reference direction in the configuration space.

Now, we calculate the following integral in (\ref{I}),%
\begin{align}
I_{lm}^{\pm}(\lambda_{z})  &  =\int_{0}^{\pi}\Psi_{\lambda_{z}}^{\pm\ast
}(\theta)N_{lm}P_{l}^{m}(\cos\theta)\sin\theta d\theta\label{In}\\
&  =\frac{N_{lm}}{2\sqrt{\pi}}\left(  \int_{0}^{\pi/2}\frac{1}{\sqrt
{\cos\theta}}\exp(i\lambda_{z}\ln\tan\theta)P_{l}^{m}(\cos\theta
)d\theta\right. \nonumber\\
&  \pm\left.  \int_{\pi/2}^{\pi}\frac{1}{\sqrt{\left\vert \cos\theta
\right\vert }}\exp(i\lambda_{z}\ln\left\vert \tan\theta\right\vert )P_{l}%
^{m}(\cos\theta)d\theta\right)  . \label{Int}%
\end{align}
With help of the variable transformation $\theta\rightarrow\pi-\theta$ in
following integral,%
\begin{align}
\int_{\pi/2}^{\pi}\frac{1}{\sqrt{\left\vert \cos\theta\right\vert }}%
\exp(i\lambda_{z}\ln\left\vert \tan\theta\right\vert )P_{l}^{m}(\cos
\theta)d\theta &  =-\int_{\pi/2}^{0}\frac{1}{\sqrt{\cos\theta}}\exp
(i\lambda_{z}\ln\tan\theta)P_{l}^{m}(-\cos\theta)d\theta\nonumber\\
&  =\int_{0}^{\pi/2}\frac{1}{\sqrt{\cos\theta}}\exp(i\lambda_{z}\ln\tan
\theta)P_{l}^{m}(-\cos\theta)d\theta, \label{intg}%
\end{align}
we have therefore for $I_{lm}^{\pm}(\lambda_{z}),$%
\begin{align}
I_{lm}^{\pm}(\lambda_{z})  &  =\frac{N_{lm}}{2\sqrt{\pi}}\int_{0}^{\pi/2}%
\exp(i\lambda_{z}\ln\tan\theta)\left[  \frac{P_{l}^{m}(\cos\theta)\pm
P_{l}^{m}(-\cos\theta)}{\sqrt{\cos\theta}}\right]  d\theta\nonumber\\
&  =\frac{N_{lm}}{2\sqrt{\pi}}\int_{0}^{\pi/2}\exp(i\lambda_{z}\ln\tan
\theta)\left[  P_{l}^{m}(\cos\theta)\pm P_{l}^{m}(-\cos\theta)\right]
\sqrt{\cos\theta}\sin\theta d\ln\tan\theta\nonumber\\
&  =\left[  1\pm(-1)^{l+m}\right]  \frac{N_{lm}}{\sqrt{2}}\int_{-\infty
}^{\infty}\frac{\exp(i\lambda_{z}u)}{\sqrt{2\pi}}P_{l}^{m}(\frac{1}%
{\sqrt{1+e^{2u}}})\frac{e^{u}}{\left(  1+e^{2u}\right)  ^{3/4}}du.
\label{Integr}%
\end{align}
In the last line we used a relation $P_{l}^{m}(-x)=$ $(-1)^{l+m}P_{l}^{m}(x)$
and introduced the variable transformation
\begin{equation}
\ln\tan\theta\rightarrow u\text{, or }\theta\rightarrow\arctan(e^{u}),\text{
}(u\in(-\infty,\infty)).
\end{equation}
The result (\ref{Integr}) is nothing but a Fourier transform of following
function,
\begin{equation}
\frac{N_{lm}}{\sqrt{2}}\left[  1\pm(-1)^{l+m}\right]  P_{l}^{m}(\frac{1}%
{\sqrt{1+e^{2u}}})\frac{e^{u}}{\left(  1+e^{2u}\right)  ^{3/4}}.
\end{equation}

Since the associated Legendre polynomial $P_{l}^{m}(x)$ (\ref{alp}) are even
or odd functions respectively corresponding to $l+m$ being even or not, the
integral $I_{lm}^{+}(\lambda_{z})=0$ with odd $l+m$ and \textit{vice versa}.
Moreover, since for given quantum numbers ($l,m$) the $P_{l}^{-m}(x)$ and
$P_{l}^{m}(x)$ differ by a factor of constant, the posmom distribution
densities for the pair of $P_{l}^{-m}(x)$ and $P_{l}^{m}(x)$ are the same. The
explicit forms $I_{lm}^{\pm}(\lambda_{z})$ for the first six nontrivial
results for $P_{l}^{m}(x)$ are with $(l,m)=(0,0)$; $l=1,m=0,1$; and
$l=2,m=0,1,2$,%
\begin{align}
I_{00}^{+}(\lambda_{z})  &  =\frac{1}{\sqrt{2\pi}}\left\{  \frac{2i}%
{2\lambda_{z}+i}\,F\left(  \frac{3}{4},\frac{1}{4}(1-2i\lambda_{z});\frac
{1}{4}(5-2i\lambda_{z});-1\right)  \right. \nonumber\\
&  \left.  -\frac{i}{\lambda_{z}-i}F\left(  \frac{3}{4},\frac{1}{2}%
(i\lambda_{z}+1);\frac{1}{2}(i\lambda_{z}+3);-1\right)  \right\}  ,
\end{align}%
\begin{align}
I_{10}^{-}(\lambda_{z})  &  =\frac{1}{4\lambda_{z}^{2}+2i\lambda_{z}+6}%
\sqrt{\frac{3}{2\pi}}\left\{  -\left(  8\lambda_{z}^{2}+2i\lambda
_{z}+15\right)  \,F\left(  \frac{1}{4},\frac{1}{2}(i\lambda_{z}+1);\frac{1}%
{2}(i\lambda_{z}+3);-1\right)  \right. \nonumber\\
&  +(15+4\lambda_{z}(\lambda_{z}-i))\,F\left(  -\frac{3}{4},\frac{1}%
{2}(i\lambda_{z}+1);\frac{1}{2}(i\lambda_{z}+3);-1\right) \nonumber\\
&  +4(3+\lambda_{z}(\lambda_{z}+2i))\,F\left(  -\frac{3}{4},\frac{1}%
{4}(3-2i\lambda_{z});\frac{1}{4}(7-2i\lambda_{z});-1\right) \nonumber\\
&  \left.  -2(7+\lambda_{z}(4\lambda_{z}+3i))\,F\left(  \frac{1}{4},\frac
{1}{4}(3-2i\lambda_{z});\frac{1}{4}(7-2i\lambda_{z});-1\right)  \right\}  ,
\end{align}%
\begin{align}
I_{11}^{+}(\lambda_{z})  &  =\frac{1}{8}\sqrt{\frac{3}{\pi}}\left\{  \frac
{1}{2\lambda_{z}+i}4(4\lambda_{z}+3i)\,F\left(  \frac{1}{4},\frac{1}%
{4}(1-2i\lambda_{z});\frac{1}{4}(5-2i\lambda_{z});-1\right)  \right.
\nonumber\\
&  -\frac{1}{2\lambda_{z}+i}8(\lambda_{z}+2i)\,F\left(  -\frac{3}{4},\frac
{1}{4}(1-2i\lambda_{z});\frac{1}{4}(5-2i\lambda_{z});-1\right) \nonumber\\
&  \left.  -e^{\frac{\pi\lambda_{z}}{2}}(9+4i\lambda_{z})B_{-1}\left(
\frac{i\lambda_{z}}{2}+1,\frac{3}{4}\right)  +e^{\frac{\pi\lambda_{z}}{2}%
}(7+2i\lambda_{z})B_{-1}\left(  \frac{i\lambda_{z}}{2}+1,\frac{7}{4}\right)
\right\}  ,
\end{align}%
\begin{align}
I_{20}^{+}(\lambda_{z})  &  =-\frac{1}{24}\sqrt{\frac{5}{2\pi}}\left\{
\frac{1}{\lambda_{z}-i}6(-2\lambda_{z}+3i)\,_{2}F_{1}\left(  -\frac{1}%
{4},\frac{1}{2}(i\lambda_{z}+1);\frac{1}{2}(i\lambda_{z}+3);-1\right)  \right.
\nonumber\\
&  +\frac{1}{\lambda_{z}-i}6(4\lambda_{z}-i)\,_{2}F_{1}\left(  \frac{3}%
{4},\frac{1}{2}(i\lambda_{z}+1);\frac{1}{2}(i\lambda_{z}+3);-1\right)
\nonumber\\
&  +\frac{\Gamma\left(  \frac{1}{4}-\frac{i\lambda_{z}}{2}\right)  }%
{\Gamma\left(  \frac{1}{4}(5-2i\lambda_{z})\right)  }6(1-i\lambda
_{z})\,F\left(  -\frac{1}{4},\frac{1}{4}(1-2i\lambda_{z});\frac{1}%
{4}(5-2i\lambda_{z});-1\right) \nonumber\\
&  +\left.  \frac{\Gamma\left(  \frac{1}{4}-\frac{i\lambda_{z}}{2}\right)
}{\Gamma\left(  \frac{1}{4}(5-2i\lambda_{z})\right)  }3(-1+4i\lambda
_{z})F\left(  \frac{3}{4},\frac{1}{4}(1-2i\lambda_{z});\frac{1}{4}%
(5-2i\lambda_{z});-1\right)  \right\}  ,
\end{align}%
\begin{align}
I_{21}^{-}(\lambda_{z})  &  =\frac{1}{8}\sqrt{\frac{5}{3\pi}}\left\{
\frac{4(4\lambda_{z}+i)}{2\lambda_{z}+3}\,F\left(  \frac{3}{4},\frac{1}%
{4}(3-2i\lambda_{z});\frac{1}{4}(7-2i\lambda_{z});-1\right)  \right.
\nonumber\\
&  -\frac{8(\lambda_{z}+2i)\,}{2\lambda_{z}+3}F\left(  -\frac{1}{4},\frac
{1}{4}(3-2i\lambda_{z});\frac{1}{4}(7-2i\lambda_{z});-1\right) \nonumber\\
&  -e^{\frac{\pi\lambda_{z}}{2}}(3+4i\lambda_{z})B_{-1}\left(  \frac
{i\lambda_{z}}{2}+1,\frac{1}{4}\right) \nonumber\\
&  +\left.  e^{\frac{\pi\lambda_{z}}{2}}(5+2i\lambda_{z})B_{-1}\left(
\frac{i\lambda_{z}}{2}+1,\frac{5}{4}\right)  \right\}  ,
\end{align}%
\begin{align}
I_{22}^{+}(\lambda_{z})  &  =\frac{1}{8(3+\lambda_{z}(2\lambda_{z}-5i))}%
\sqrt{\frac{5}{3\pi}}\left\{  (7+4\lambda_{z}(\lambda_{z}-3i))\,F\left(
-\frac{1}{4},\frac{1}{2}(i\lambda_{z}+3);\frac{1}{2}(i\lambda_{z}%
+5);-1\right)  \right. \nonumber\\
&  +4(3+\lambda_{z}(\lambda_{z}-2i))\,F\left(  -\frac{1}{4},\frac{1}%
{4}(1-2i\lambda_{z});\frac{1}{4}(5-2i\lambda_{z});-1\right) \nonumber\\
&  +(-7+2(-4\lambda_{z}+5i)\lambda_{z})\,F\left(  \frac{3}{4},\frac{1}%
{2}(i\lambda_{z}+3);\frac{1}{2}(i\lambda_{z}+5);-1\right) \nonumber\\
&  +\left.  2(9+(-4\lambda_{z}+15i)\lambda_{z})\,F\left(  \frac{3}{4},\frac
{1}{4}(1-2i\lambda_{z});\frac{1}{4}(5-2i\lambda_{z});-1\right)  \right\}  ,
\end{align}
where $F(a,b;c;x)$ is the hypergeometric function, and $B_{z}(a,b)$ are
incomplete Beta function of rank $z$: $B_{z}(a,b)=z^{a}\sum_{n=0}^{\infty
}z^{n}\Gamma(n+1-b)/(\Gamma(1-b)n!(a+n))$. For an arbitrary $I_{lm}^{\pm
}(\lambda_{z})$ with larger ($l,m$), we observe that the expression has in
general four terms but each of the coefficients before the\ hypergeometric
functions becomes a ratio of two polynomials of $\lambda_{z}$ with highest
terms being not exceeding $(\lambda_{z})^{l}$.

The probability distributions $\left\vert I_{lm}(\lambda_{z})\right\vert ^{2}$
for rotational states represented by spherical harmonics $Y_{lm}%
(\theta,\varphi)$ are plotted for cases $l=0$, $l=3$, and $l=20$ in Figures 1,
2, and 3, respectively. On the whole, they bear a striking resemblance to the
momentum distributions of stationary states for the one-dimensional simple
harmonic oscillator. It is perfectly understandable that from the force
operator ${\dot{p}}_{{i}}\equiv$ $[{p}_{{i}}{,H}]/(i\hbar)=-\{x_{i}/r,H\}$
$\sim-x_{i}$ with ${H}$ being the rotational Hamiltonian, and\ $\{U,V\}\equiv
UV+VU$. In other words, for a classical state, the force is restoring and
proportional to the displacement, and the quantity $x_{i}p_{i}$ has thus a
half period as $x_{i}$ or $p_{i}$ has.


\begin{figure}
[htb]%
\includegraphics[width=0.9\textwidth]{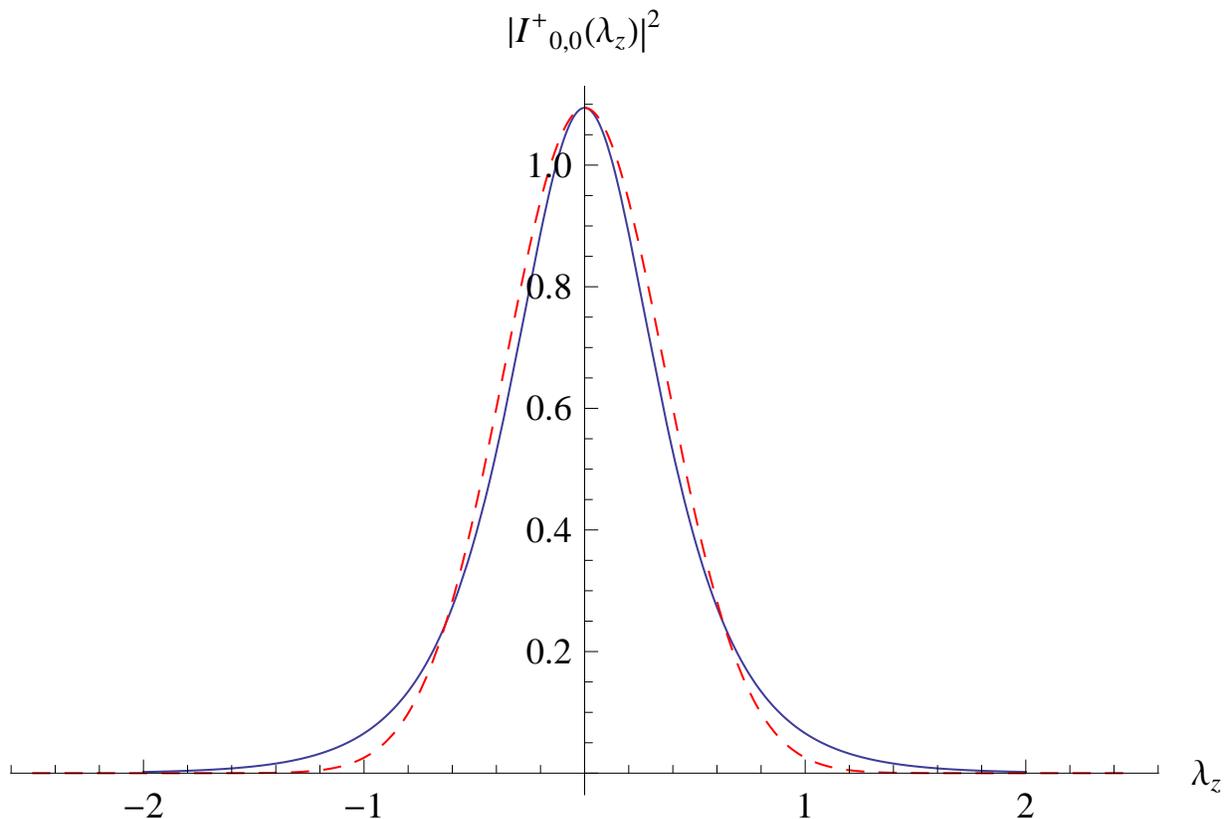}\caption{Distribution density of $(x_{i}p_{x_{i}}+p_{x_{i}}x_{i})/2$ for the ground
state of ground rotational state $Y_{0,0}=1/(\sqrt {4\pi})$ (solid line), and the momentum distribution density for the ground
state of one-dimensional simple harmonic oscillator (dashed line). They are
almost identical. In all figures, the posmoms and the momenta are made dimensionless.}\label{figure 1}%

\end{figure}

\begin{figure}
[htb]%
\includegraphics[width=0.9\textwidth]{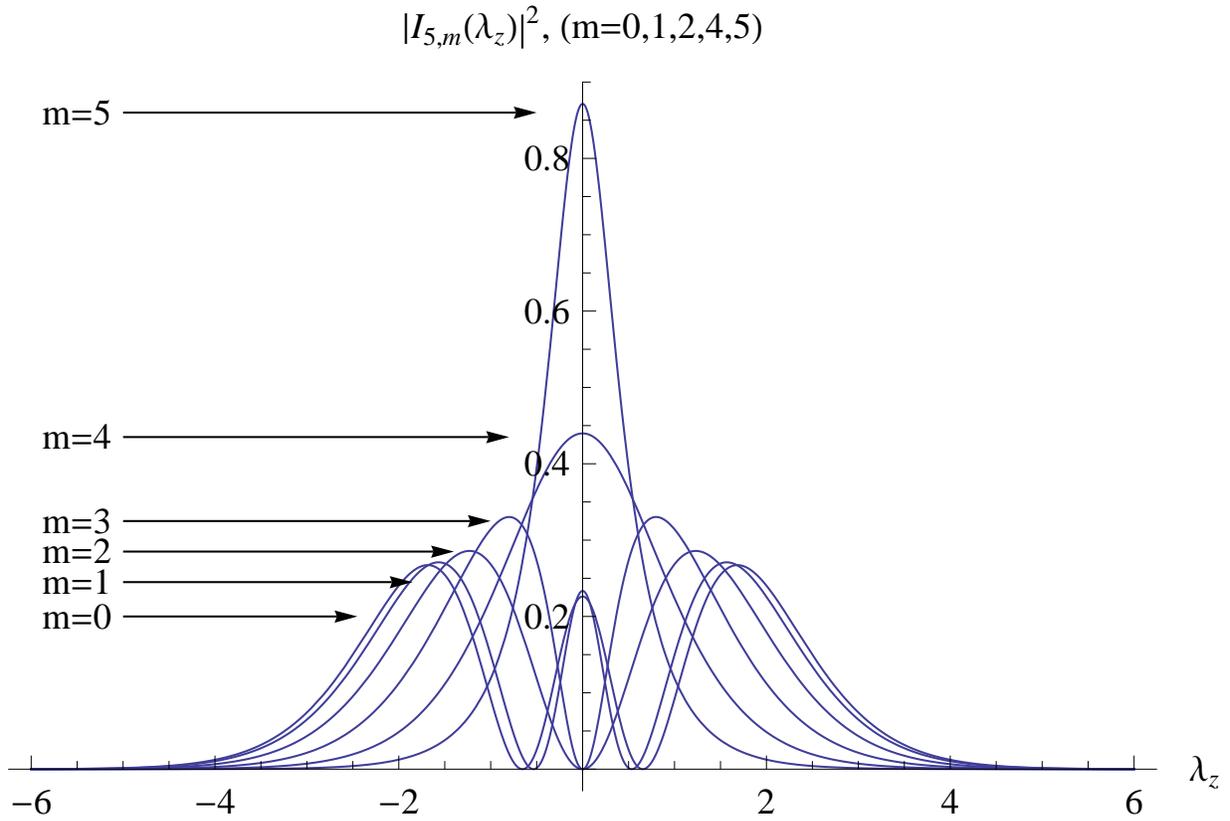}\caption{Distribution density of $(x_{i}p_{x_{i}}+p_{x_{i}}x_{i})/2$ for the rotational states $Y_{lm}(\theta,\varphi)$ with $l=5$ and
$m=0,1,2,3,4,5$, they have number of nodes $=2,2,1,1,0,0$ respectively. It is worthy of stressing that for a given set ($l,m$),
i.e. each curve in this figure, behaves like a stationary harmonic oscillator state.}\label{figure 2}%

\end{figure}

\begin{figure}
[htb]%
\includegraphics[width=0.9\textwidth]{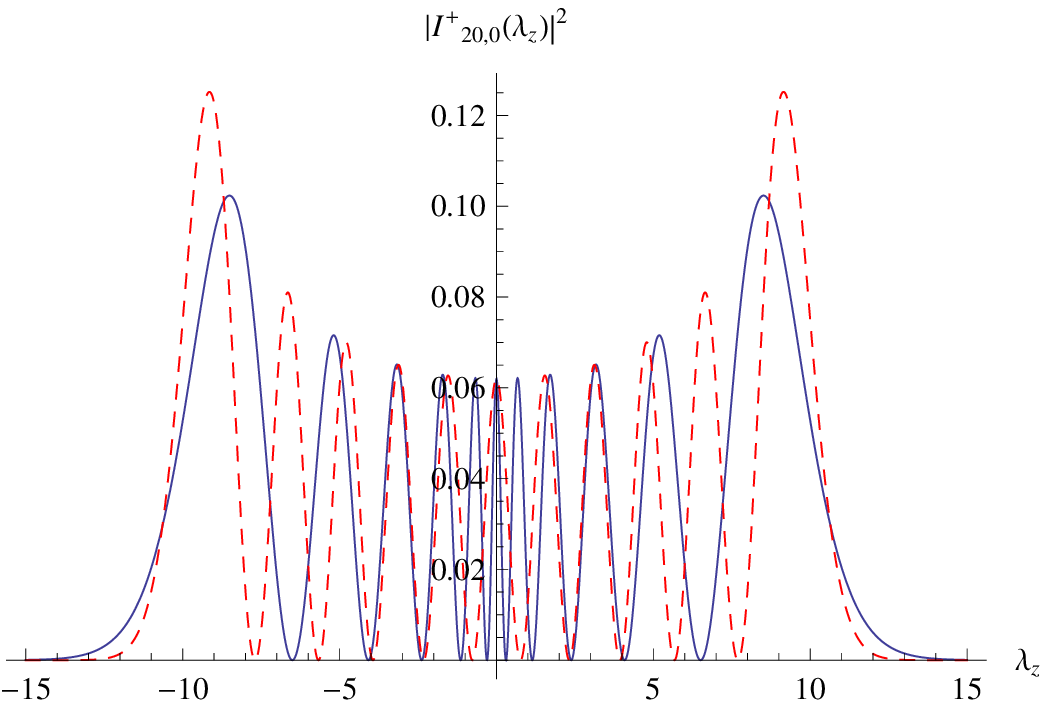}\caption{Distribution density of $(x_{i}p_{x_{i}}+p_{x_{i}}x_{i})/2$ for rotational state
$Y_{20,0}(\theta,\varphi)$ (solid line), and the
momentum distribution density for the $10$th excited state of one-dimensional
simple harmonic oscillator (dashed line). Since both probabilities in a small
interval $\Delta \lambda_{z}$ are similar, they have the same classical
limit: the simple harmonic oscillator. }\label{figure 3}
\end{figure}


\section{Conclusions}

The posmom ${Q}_{i}$ offers a potential new way to understand the quantum
motions of an atom and a molecule. This study explores the posmom on
two-dimensional surface, and identify that the momentum in it is the geometric
momentum that is recently proposed to properly describe of the momentum for
the constrained motions. From the commutation relations $[{Q}_{i},{L}_{{i}%
}]=0$, $(i=1,2,3)$, we have three complete sets of commuting observables, and
they are equivalent with each other upon a rotation of coordinates. Thus a
novel dynamical representation based on two observables, (${Q}_{z},L_{z}$) in
the present paper, is successfully constructed, and any states on the
two-dimensional surface can go through a posmometry analysis. Because the free
rotation is ubiquitous in microscopic domain, we carry out the posmom
distributions of spherical harmonics. Results show that the posmometry bears a
striking resemblance to the momentum distributions of stationary states for
the one-dimensional simple harmonic oscillators, therefore riches our
appreciation of the quantum dynamical behavior.

\begin{acknowledgments}
This work is financially supported by National Natural Science Foundation of
China under Grant No. 11175063
\end{acknowledgments}

\end{document}